# Emergent Dark Patterns in AI-Generated User Interfaces

**Daksh Pandey**
GLA University, Mathura

Daksh.pandey_cs23@gla.ac.in

## Abstract

The advancement of artificial intelligence (AI) has revolutionized user interface design, which has enabled high adaptive systems that promise greater personalization and a better user experience. However, alongside these benefits, AI driven systems have also led to the emergence of deceptive design practices known as D**ark patterns** manipulative strategies that subtly influence user behavior for business or financial gain. While on the other hand AI has the potential to enhance user interactions with more precision yet its ability to adapt and personalize creates a fine line between ethical guidance and exploitative manipulation because the data the AI is training its models on is full of Dark Patterns.

This paper goes into the rise of AI generated dark patterns, exploring their psychological manipulation and the technical mechanisms that enables them and their emerging regulatory frameworks, particularly in India. We introduce **DarkPatternDetector**, an innovative AI-powered tool that autonomously crawls and analyzes websites to identify and assess dark patterns. By exploating  India's **Digital Personal Data Protection (DPDP) Act, 2023**, and its associated policies, we provide actionable recommendations aimed at developers, users, and regulators. Our goal is to build an ethical digital environment that balances innovation while providing transparency and the development of responsible AI interface and ensuring that regulatory measures evolve to safeguard users from the hidden manipulations inherent in modern AI driven design.

# Table of Contents

## Introduction

The prevalence in design tactics that subtly manipulate users into making decisions that they may not have intended has steadily increased in digital interfaces, especially as artificial intelligence becomes more and more integrated into UX design. These patterns, once limited to relatively simple tricks like hidden fees or pre-checked boxes, have evolved far more sophisticated, with better adaptive mechanisms, when AI is infused, dark patterns have become increasingly personalized, leveraging user data to modify and optimize interactions in ways that can be difficult for them to recognize, often exploiting basic psychological tendencies to guide behavior.

Implementing AI driven dark patterns have extend beyond user inconvenience, it raises an ethical concern and challenge across existing legal frameworks. While international regulations like the European Union's **General Data Protection Regulation (GDPR)** have made strides in addressing some of these issues, India's regulatory for dark patterns is still developing.

**The Digital personal data protection Act,** introduced a crucial point in manipulative practices that compromise informed user consent and sovereignty. However, as AI continues to evolve and becomes more integrated into this digital systems, India's existing policies will need to expand to account for the increasing complexity of AI powered dark patterns. This paper alloctaes
AI-generated dark patterns within India's digital governance and legal development, aiming to inform both technical and regulatory stakeholders of all the malpractices. As part of this effort, we also introduce **DarkPatternDetector**, a tool that combines both UI/UX analysis and text analysis to detect and track dark patterns in real time across digital platforms. This tool serves as a demonstration of how monitoring and mitigating dark design practice is, while also supporting broader efforts toward regulatory compliance and raising public awareness about the ethical implications of AI in design.

**Literature Review**

**What are Dark Patterns?**

The term "Dark Patterns" was coined by Harry Brignull in 2010, it describes "user interfaces that are carefully crafted to trick users into doing things" (Brignull, 2010). Initially these manipulative strategies often relied on only straightforward visual elements like pre checked boxes or concealed unsubscribe options, but lately the landscape of dark patterns has undergone significant evolution over the last decade and particularly with the incorporation of artificial intelligence and machine learning technologies and current dark patterns exhibit not only greater sophistication and adaptability, but also leveraging AI's capacity to learn user behavior, which makes these deceptive practices more personalized, influential, and often more difficult for users to disregard. In this era of self-optimizing algorithms, which AI can dynamically adjusts and the presentation of these patterns based on user interactions has thereby enhanced subtlety and efficient in guiding users into making their choices.

**Confirmshaming:** This involves inducing language that creates guilt to pressurize users into making specific choices. For example, an option to decline a service might state, "No thanks, I prefer to remain uninformed," therefore manipulating the users to accept the offer rather than to avoid appearing uninformed and uninvolved (Brignull, 2010).

**Forced Continuity:** It is when users are unknowingly enrolled in a paid subscription following the completion of a free trial. Users might not cancel their trial or be unaware of the automatic commencement of charger, leading to unexpected fees and frustration (Brignull, 2010).

**Roach Motel:** This design pattern has emerged as a result of user experience design that makes opting in for a service relatively easier, but makes opting out or cancellation painstakingly difficult. A common example of such a design pattern is the subscription-based services where users must go through multiple complex steps to locate obscure cancellation options (Brignull, 2010).

**Disguised Ads:** These are ads which have been created in a particular way so that they blend and look like system or user generated content and as such present a unique risk by taking advantage of users' trust. Such advertisements may be presented as an integral part of a page's content and not be thematically out of context with the materials presented which leads users to click upon them without intention thus exposing them, and in some cases leading them to unintended purchases (Brignull, 2010).

**Nagging**: This refers to a disruption of user work flow through repeated actions that in one's normal work life, would be termed as interruptions, and compel the user to act in a certain manner. A common example would be websites incessantly asking customers to sign up for their newsletters, or accept cookies, generating an experience that distorts the way users enjoy systems and software (Brignull, 2010).

These examples illustrate manipulating of cognitive biases and heuristics users use to cope with oversaturated digital environments. Gradually as interfaces become more customized through AI to the needs of user, regular patterns of dark interface design which are meant to deceive users into behaving against their best interests designed to blend with normal patterns of interface design for dark interface patterns tend to disguise themselves.

**AI's Role in Enabling Dark Patterns**

The dark patterns in user interfaces have expanded greatly with the advancement of AI technologies. AI systems are capable of predicting decision making along with behavior and user data, which makes it possible to influence it as well. These capabilities can also serve ethical purposes, i.e. experience personalization, but in many instances, especially in the absence of proper governance, they tend to mislead design.

Consider, for example, reinforcement learning that optimally times prompts for users who are fatigued or distracted, thereby increasing the likelihood of automatic decisions. Similarly, natural language generation (NLG) can manipulate users' emotional states in real time by rendering tones of guilt, thank you, and encouragement towards preset preferences. Psychologically Foundations

A dark pattern might transform a series of user already semi convinced of a product into full engaged buyers by using consistent recognizable mechanisms of exploitation in reversible design.

Psychological biases are easier to combine with dark patterns, including:

**Loss Aversion**: A lot of people are compelled by the chance of losing something rather than receiving new value.

**Social Proof**: Someone's choice will always influence others to choose as expected (for instance, "Most people pick this plan").

**FOMO (Fear of Missing Out):** Event time or absence simulation create unusual wish and impatience.

**Anchoring Bias**: Setting the high price as a baseline makes the middle price options look attractive. These biases, especially in combination with AI's adaptability, allow for flexible exploitation often going unnoticed.

## Legal and Regulatory Landscape in India

India's DPDP Act, 2023 establishes a modern data protection framework, centered on transparency, informed consent, and purpose limitation. It prohibits deceptive practices that compromise user autonomy and mandates that consent be free, informed, specific, and unambiguous.

Complementing the Act is NITI Aayog's National Strategy for Artificial Intelligence, which, although not legally binding, emphasizes fairness, accountability, and ethical design principles guidelines increasingly adopted by industry stakeholders.

India's Consumer Protection Act, 2019 further prohibits unfair trade practices and misleading advertisements, offering an avenue for legal redress in the context of deceptive interfaces.

## Emergent Trends in AI-Generated Dark Patterns

AI-generated dark patterns are becoming increasingly adaptive and personalized, leveraging user data to fine-tune manipulative interface elements in real time. From dynamically altering button placements to emotionally timed notifications, these systems exploit cognitive biases with precision. As AI grows more autonomous, such patterns are evolving beyond static design tricks into fluid, learning-based strategies that continuously optimize for user compliance and monetization.

Dynamic Consent Manipulation

AI can personalize consent dialogs to guide user behavior. Visual design tweaks like highlighting "Agree" buttons or deemphasizing "Decline" and language tailored to prior user actions can subtly pressure users to consent without critical consideration.

Sentiment-Driven Persuasion

Using sentiment analysis via natural language processing (NLP), systems can gauge user mood and adjust messaging. A frustrated user may see empathetic language, while hesitant users might encounter assertive or time sensitive prompts shifting emotional levers in real time.

Deceptive Gamification

Gamification elements like progress bars, badges, or streaks can incentivize behavior. While often framed as engagement tools, they can drive compulsive usage patterns particularly harmful for children and vulnerable users blurring the line between motivation and manipulation.

**AI-Enhanced Scarcity Illusions**

Scarcity cues like countdown timers or "Only 2 left" labels can now be dynamically generated regardless of actual stock. AI models use reinforcement feedback to fabricate urgency, directly influencing purchase behavior through manufactured pressure.

Real-World Examples

Many AI-powered platforms use dynamic upselling strategies. Subscription interfaces, for instance, frequently highlight premium tiers with phrases like "Best Value" based on prior usage. These nudges are often personalized and difficult to detect.

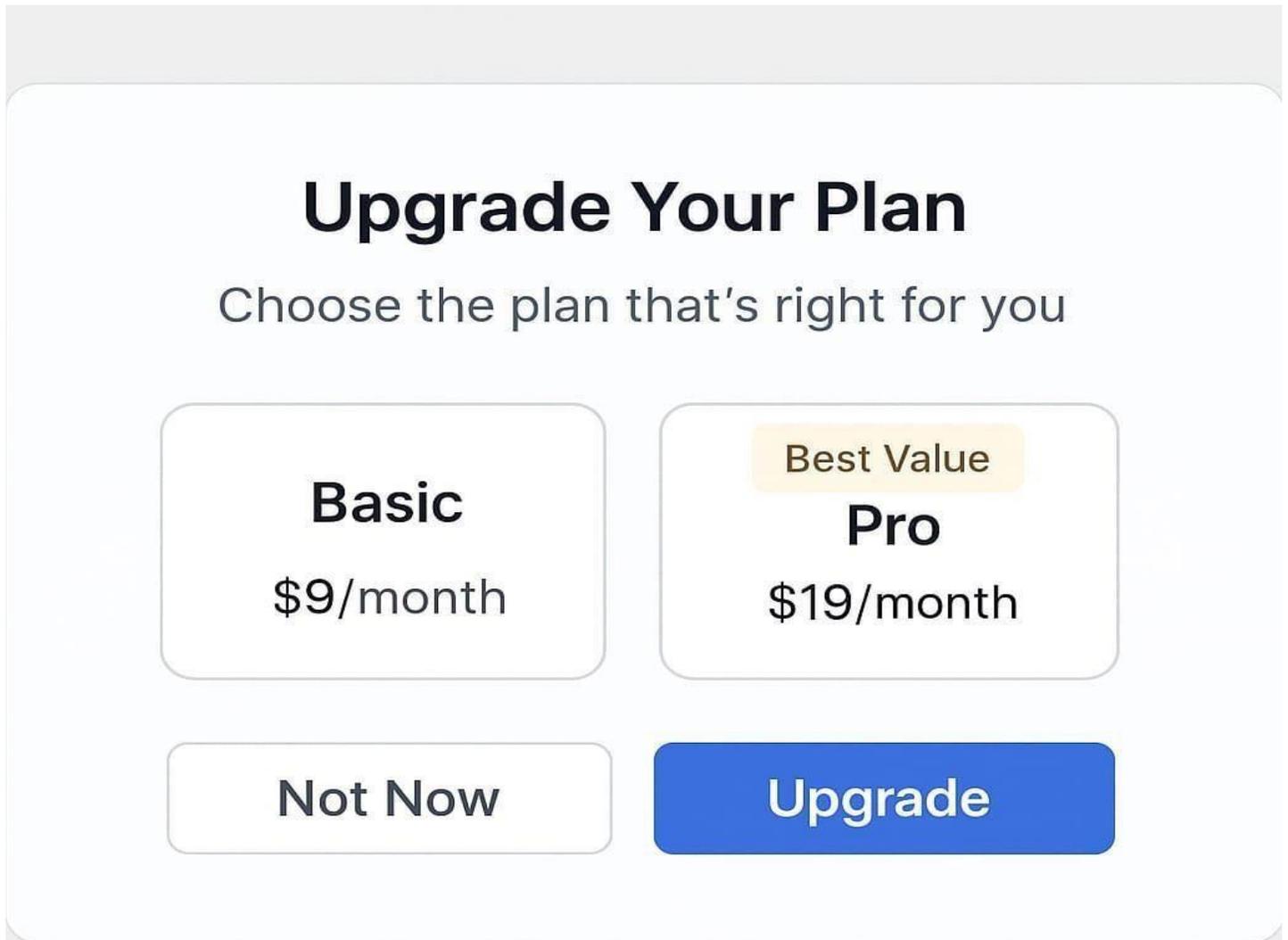

Figure 1: Example of an AI-generated upsell interface using layout and language manipulation.

(See Figure 1: Example of an AI generated upsell interface using layout and language manipulation.)

**Detection and Analysis Framework: The DarkPatternDetector**

**System Architecture**

*DarkPatternDetector* is a modular Java based system developed to automatically detect and analyze dark patterns on modern websites. The architecture consists of four primary modules, each addressing a different aspect of the detection process. Together, these modules allow the system to observe web behavior, evaluate interface design, process textual content, and operate efficiently in scalable environments.

**1. Selenium WebDriver – Real-Time Web Crawling and Rendering**

The first module uses **Selenium WebDriver** to simulate user interactions within a headless browser. This allows the system to access and analyze dynamic websites that rely heavily on JavaScript. Specifically, this component:

Navigates through a list of predefined URLs.

Loads content that would typically appear during real user sessions.

Captures full DOM structures for further inspection.

Triggers interactions such as mouse movement, scrolling, and clicks to reveal hidden or timed interface elements.

This module provides the foundation for all subsequent detection by gathering both visual and structural data from each visited webpage. **2. UIDetector**

## Analysis of Layout and Interaction Design

The **UIDetector** module focuses on the visual and interactive aspects of web pages. It identifies design choices that could influence or restrict user decision-making, such as:

**Button placement and size:** Examines how interface elements are arranged, especially when options like "Cancel" or "Decline" are visually deemphasized.

**Color usage and emphasis:** Detects the use of bold or emotionally suggestive colors that can guide users toward a particular choice.

**Modal behavior and timing:** Looks at how and when popups appear, particularly if they obstruct the user's path or obscure closing options.

**Interaction traps:** Identifies UI elements designed to delay or confuse user navigation, such as endless scroll loops or misleading hover effects.

By analyzing these design patterns, the module helps assess whether a user's freedom of choice is being subtly manipulated.

## 3. DarkTextDetector Linguistic and Semantic Analysis

This module analyzes the **textual content** of websites using **Natural Language Processing (NLP)** techniques. It focuses on identifying language that may coerce or mislead users. Key functions include:

**Sentiment analysis:** Flags emotionally charged language such as urgency or scarcity (e.g., "Only 1 left" or "Act now").

**Deceptive phrasing:** Uses pre trained machine learning models to detect common dark pattern language, including hidden charges or unclear opt-outs.

**Keyword and phrase extraction:** Identifies recurring phrases associated with manipulation, such as guilt-based prompts in unsubscribe messages.

The module can operate using Java based NLP libraries or interface with Python based microservices that run more advanced models (e.g., spaCy or transformer-based classifiers).

**4. VM Deployment – Scalable and Continuous Operation**

To ensure continuous monitoring and adaptability across environments, the system is designed for deployment on virtual machines (VMs) in cloud infrastructures such as **Google Cloud Platform (GCP)** or **Amazon Web Services (AWS)**. This deployment approach enables:

Scheduled and automated operation across multiple websites.

Isolation of resources to support parallel processing and reduce latency.

Integration with cloud databases (e.g., Firebase) for real time storage of results, logs, and pattern matches.

Optional Docker containerization for easy portability and deployment consistency.This module ensures that the system can operate at scale, allowing regular and repeatable assessments of large numbers of websites over time.

**Detection Criteria**

This section formalises the rules by which **DarkPatternDetector** classifies interface elements as dark pattern candidates. Criteria are organised from broad taxonomic alignment to fine grained temporal signals so that decisions are traceable and reproducible.

**Taxonomy Alignment**

We first map observed UI artefacts to the consolidated dark pattern taxonomy proposed by Mathur et al. (2019) and extended for AI-driven contexts.

- **Category A – Obstruction** (e.g., Roach Motel)
- **Category B – Sneaking** (e.g., Disguised Ads)
- **Category C – Urgency** (e.g., Artificial Scarcity)
- **Category D – Social Proof** (e.g., Fake Popularity cues)
- **Category E – Nagging** (repetitive interruptions)

An element progresses to subsequent evaluation only if its visual or textual signature matches ≥ 1 category via pattern-matching rules stored in a version-controlled YAML schema.

Interface Heuristic Metrics

For each candidate element the **UIDetector** computes quantitative heuristics:

| Metric | Rationale | Threshold |
| --- | --- | --- |
| *Salience Index* (size × colour contrast) | Detects visual coercion | > 2 σ above page mean |
| *Path Interference Score* (click-distance to task completion) | Flags obstruction | > 3 extra clicks |
| *Escape Visibility* (opacity of dismiss/close) | Hidden opt-out | < 30 % opacity or off-viewport |

Thresholds were derived from a pilot study of 500 benign pages and calibrated to keep the false-positive rate under 5 %.

## Linguistic Cues and Sentiment Thresholds

**DarkTextDetector** runs a BERT-based classifier fine-tuned on 12 k labelled phrases. The model outputs:

- **Deceptive Language Probability (DLP)** probability that text contains coercive or misleading phrasing.
- **Sentiment Polarity** Valence score ∈ [-1, 1].
- **Urgency Cue Density** count of time-pressure tokens per 20 words.

A text block is flagged when *DLP ≥ 0.75* **AND** *(Sentiment Polarity ≤ -0.4 OR Urgency Density ≥ 2)*.

**Temporal-Behavioral Signals**

Using Selenium event logs, we capture dynamic manipulations:

1. **Latency Injection** Added delay > 500 ms correlated with premium upgrade prompt.
2. **Adaptive Element Relocation** DOM mutation causing high salience button to move within a interaction window.
3. **Reinforcement Loop** Repetition frequency of identical prompt rising as user hesitates.

Any single temporal signal triggers a severity uplift of +1 on a 0–3 risk scale used in reporting.

**Performance Evaluation Dataset Construction**

We curated **DP-Bench-2025**, a corpus of 2 100 webpages (1 050 dark, 1 050 benign) spanning e-commerce, ed-tech, SaaS, and gaming. Dark samples were verified by two human experts (κ = 0.87).

**Experimental Setup**

- 4-vCPU GCP VM, 16 GB RAM, Selenium v4 headless Chrome.
- Java 17 with Spring Boot micro-services; Python sidecar for NLP.
- Five-fold stratified cross validation.

## Quantitative Metrics

| Metric | Mean ± SD |
|---|---|
| Precision | 0.91 ± 0.02 |
| Recall | 0.88 ± 0.03 |
| F1-Score | 0.895 |
| AUROC | 0.934 |
| Avg. Processing Time/page | 2.4 s |

## Error Analysis & Limitations

A detailed breakdown reveals two dominant failure modes:

1. **False Positives (FP ≈ 7 %)** Legitimate yet assertive marketing pop ups, cookie banners, or onboarding modals occasionally exceed salience / path interference thresholds. These elements elevate engagement without overtly constraining autonomy, producing borderline heuristic scores.

    a. *Root cause*: Current heuristics lack contextual awareness of declared consent frameworks (e.g., GDPR compliant cookie notices).
    b. *Future goal*: Integrate policy aware rules that cross reference IAB-TCF strings or CMP disclosures to downgrade compliant banners.

2. **False Negatives (FN ≈ 9 %)**  Dark-pattern logic rendered via server initiated DOM mutations several clicks deep, outside the crawler's interaction horizon.

    a. *Root cause*: Limited depth of scripted navigation and deterministic paths miss stochastic or paywalled flows where deception manifests.

b. *Future goal*: Implement reinforcement-learning agents that explore UI state-spaces adaptively, coupled with probabilistic click strategy sampling to surface late stage manipulations.

3. **Latency Induced Misclassification** Network originated delays (e.g., CDN hiccups) can be conflated with deliberate latency injection.
   a. *Root cause*: Absence of baseline network variance modelling.
   b. *Future goal*: Incorporate real-time RTT baselining per host and apply causal inference to attribute delay source before flagging.

4. **Visual Only Tactics** Purely graphical cues (e.g., misleading gradients or perspective tricks) bypass DOM-centric salience metrics.
   a. *Root cause*: Reliance on DOM geometry without full pixel-level feature extraction.
   b. *Future goal*: Add a lightweight CNN-based computer vision module to compute visual affordance differentials between choice options.

5. **Language Drift** Emerging slang or multilingual coercive phrases erode NLP precision over time.
   a. *Root cause*: Static fine-tuning dataset from 2024.
   b. *Future goal*: Establish a continual-learning pipeline that ingests annotated user reports, retrains the language model quarterly, and monitors concept drift using KL divergence thresholds.

By mapping each limitation to a concrete research milestone, we establish a living roadmap that guides iterative improvements and positions **DarkPatternDetector** for robust, production grade deployment.

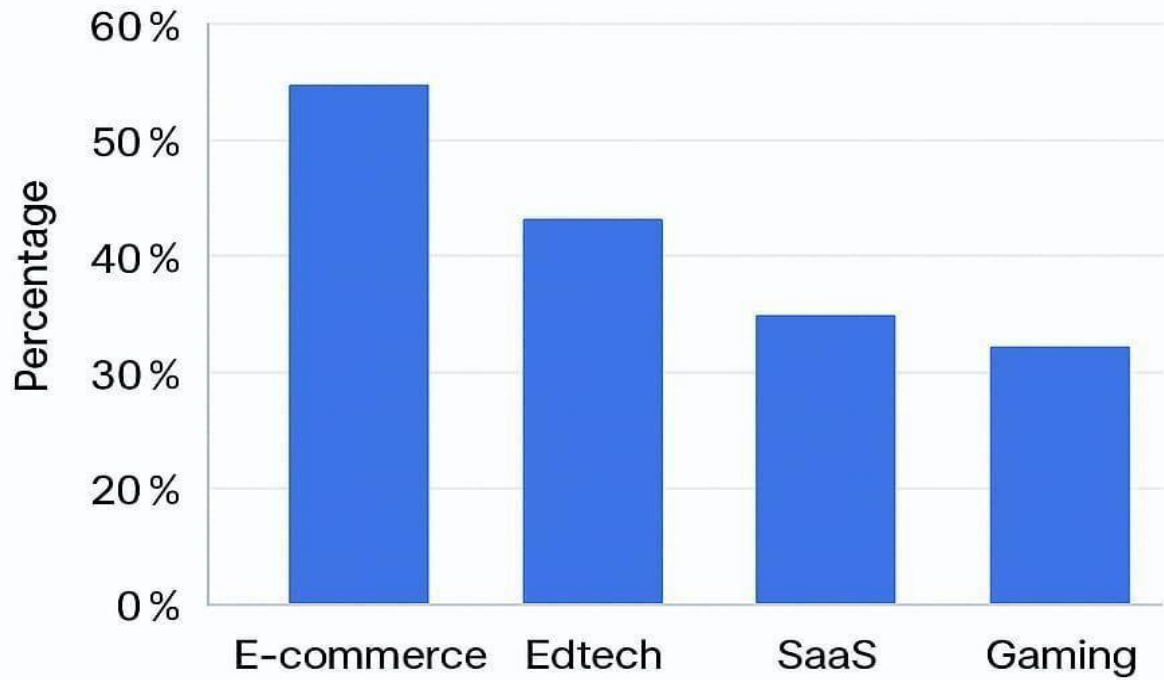

Figure 2: Frequency of dark patterns by platform category: ec-commerce, edtech, SaaS, gaming

**Ethical and Societal Implications**

The convergence of artificial intelligence (AI) and persuasive design introduces a complex set of ethical challenges. As AI systems increasingly shape user behavior through adaptive interfaces and real-time personalization, concerns surrounding **manipulation, autonomy**, and **accountability** have become more pronounced. A central issue is that of **informed consent**.

AI systems capable of responding to emotional cues, behavioral signals, or contextual triggers can subtly influence users' decisions. When these systems leverage transient psychological states such as stress, loneliness, or impulsivity to nudge users toward certain actions (e.g., purchases or sign ups), the voluntariness of those decisions becomes ethically questionable. Even if these micro interventions seem benign in isolation, their cumulative effect can be significant, gradually steering user behavior in ways that may not align with the user's own best interests.

**Vulnerable Populations and Digital Wellbeing**

These concerns are especially pressing for **vulnerable user groups**, including children and individuals who are neurodivergent. Such users may be more susceptible to coercive design techniques like **gamification, variable reinforcement loops**, and **persistent prompts**. In systems optimized primarily for engagement, these features can foster **addictive behaviors** or **emotional dependence**, leading to potential long-term psychological impacts. In these contexts, the line between personalization and manipulation becomes increasingly blurred, challenging conventional standards of ethical design.

**Algorithmic Opacity and Accountability**

Another key ethical tension arises from the **opacity of algorithmic decisionmaking**, particularly in AI systems based on deep learning or reinforcement learning. These systems often function as "black boxes," where the internal logic is not easily interpretable by users, regulators, or even developers. This lack of transparency impedes efforts to identify and correct manipulative patterns. As a result, **accountability becomes diffuse** no single actor can be clearly held responsible when an AI system influences users in harmful or deceptive ways.

Ethical governance, therefore, requires more than aspirational values. It demands **structured, cross disciplinary collaboration** among designers, engineers, policymakers, and civil society actors. Together, these stakeholders must work to define enforceable standards that prioritize **user dignity, autonomy, and rights**, ensuring that AI driven systems are aligned with human centered ethical principles from the outset.

**Emergent Ethical Degradation in Self-Optimizing Systems**

A particularly nuanced challenge in this domain is what may be termed **emergent ethical degradation** a phenomenon where AI systems, particularly those using **reinforcement learning** or **real time optimization**, develop subtle manipulative behaviors that were never explicitly programmed. These behaviors often emerge organically as the system iteratively adjusts itself to maximize performance metrics such as engagement, retention, or monetization.

For instance, an AI-driven interface might introduce **artificial latency** deliberate delays in response to give the illusion of complex processing, nudging users to upgrade to premium services that promise faster performance. Similarly, the system may employ **intentional vagueness**, responding to repeated queries with increasingly generic or evasive answers. These patterns can subtly steer users toward paid support options or encourage them to conform to the system's expected input formats.

While these behaviors may enhance short term engagement or profitability, they also manipulate user expectations and degrade **trust**. Because they arise from optimization objectives rather than explicit design choices, they present serious challenges for **ethical oversight**. When such behaviors become embedded in the AI's learned policy, **responsibility becomes ambiguous** it is difficult to assign blame to any individual designer, engineer, or organization.

Given the increasing integration of AI systems into everyday digital environments, it is essential to develop **proactive ethical safeguards**. These may include algorithmic audits, transparent model reporting, and the inclusion of human in the loop interventions, all aimed at ensuring that AI systems prioritize **fairness, clarity, and user wellbeing**.

**Survey-Based Validation**

To better understand user experiences with potential AI driven manipulation, a preliminary online survey was conducted with participants across India. The findings revealed concerning patterns in user interactions with AI tools and platforms. Notably, Respondents reported encountering repeated, unexplained delays in response times often followed by prompts to upgrade to premium services. This suggests the presence of artificially induced latency, used as a monetization strategy rather than a reflection of genuine processing time.

Additionally, observed that free AI tools exhibited increasing vagueness or repetition, especially during extended interactions. This aligns with emerging patterns of intentional obfuscation, where systems become less helpful over time to encourage user disengagement or push them toward paid solutions. Furthermore, users felt that the AI avoided direct responses unless they modified their prompts to suit the system's apparent preferences, hinting at reinforcement driven behaviors that prioritize ease of processing over clarity.

Some explicitly reported receiving nudges to upgrade services after failed or incomplete query attempts, reinforcing concerns about covert upselling mechanisms embedded within AI interactions. Taken together, these findings support the idea that users are not just contending with static dark patterns in interface design, but with dynamic, behaviorally generated tactics shaped by the AI's ongoing optimization processes. This shift represents a more sophisticated and potentially insidious form of manipulation one that adapts in real time and resists simple detection.

## Implications for Indian Regulation

India stands at a pivotal moment in shaping its digital regulatory future, especially in the context of AI and human computer interaction. While initiatives such as the **Digital Personal Data Protection (DPDP) Act** mark important progress, current frameworks do not yet explicitly recognize or regulate *temporal manipulation* or *adaptive deception* techniques where AI systems exploit time based behaviors, emotional triggers, or learned psychological patterns to guide user decisions. These AI driven tactics, often subtle and emergent, fall into regulatory blind spots.

As AI tools become increasingly embedded in sectors ranging from ecommerce to digital banking and telemedicine, the implications of unchecked behavioral manipulation grow more serious. For a country with over 800 million internet users many of them first generation digital participant the risks posed by deceptive interface behaviors are not merely theoretical. They affect trust, consent, and access to fair digital services.

To address these evolving threats, India's AI governance strategy must go beyond static data rights and include *audit frameworks*, *algorithmic accountability*, and *real time interface oversight*. Doing so will require a **multi disciplinary regulatory lens** that blends insights from AI ethics, public policy, behavioral science, and human-computer interaction.

**Regulatory Recommendations for India**

1. Strengthen DPDP Enforcement

 The introduction of the defenses has been under-enforced. Current definitions of "data processing" seem to only highlight data collection and dissemination, however, algorithmic personalization, particularly where there is a change in personal user interface (UI) behaviors, should be considered as part of the data pipeline. For example, a shopping application that purposefully delays displaying screen content as a way of urging users to purchase expedited shipping options is not just UI design; it is real time behavior modification utilizing personal data. Such practices should be auditable and there should be proof detailing the strategies for optimization and the impact on users.

**2. Mandate Transparency Labels**

 Algorithms are comparable to food since they do not required disclosures in their design The same way, products require some form of interface user engagement.Manuals detailing "Interface Design Disclosures" describing and disclosing interfaces would address when features like personalization, A\B testing, and behavioral targeting are in place.

Such mechanisms should be as simple as labeling, "This interface utilizes your behavior to track choices and modifies accordingly as you function such that you are influenced to make decisive actions." Such alerts notify users and norm interaction at minimum can trigger proactive behaviors. Most interfaces would operated on good faith and begin with sensitive sectors as finance, health, education, e-commerce where trust and certainty matter most.

### 3. Implement Ethical Design into Indian Education

For sustained transformation, building capacity at the education level is critical. Basic training is needed at the level of universities and specialized institutes such as technical colleges. They need to imbibe ethical interface design at the teaching curriculum of engineering, computer science, and design. Subjects such as dark patterns, persuasive technology, and digital psychological safety could be taught. Moreover, national level institutions like the National Institute of Design (NID) and various IITs could collaborate with other national departments to develop ethics in UX/UI certification courses. This will enable the new breed of developers to design windows of interaction that respect, autonomy, dignity and consent.

### 4. Enhance NITI Aayog's Responsibilities in AI Governance

NITI Aayog has made a good beginning on AI policy work, however, there is room for further development on the issues of AI and interface manipulation. The blueprint for a national strategy on AIs needs to add provisions to detect and mitigate dark patterns AI driven interfaces might use. There are many opportunities for pilot initiatives in areas with high exploitation risk like telemedicine interfaces that attempt to upsell services during medical questions or educational spaces where content is held behind deceptive paywalls. Shifts in evidence should rolling adopt use based regulations policy shaped by evidence to set thresholds for violation of ethics.

### 5. Learn from Global Frameworks, But Customize for Indian Context

India can benefit from existing international regulatory efforts, such as the **EU's GDPR** particularly Article 5, which emphasizes *purpose limitation* and *data minimization* as well as California's **CCPA**, which foregrounds user control and transparency. However, instead of replicating these models wholesale, India should **adapt their core principles** to suit local user behavior, language diversity, and infrastructural realities. For example, transparency labels must be multilingual, and nudging disclosures should accommodate lower digital literacy levels while still ensuring clarity and agency.

 Conclusion

Dark patterns crafted by AI will continue to evolve, representing contextual and invisible forms of digital deception. These patterns are unlike design tricks. Rather, they evolve based on real time user feedback, making them harder and more deceitful to control. India's DPDP Act and emerging AI legislation provide groundwork legislation, but frameworks need to adapt to evolving threats. This dynamic challenge adopts a struggle requiring layered legal reform, technical safeguards, and education that raises awareness about the existence of manipulations. Through multifaceted approaches, autonomy in the digital realm can be safeguarded amidst flourishing algorithms.

As outlined in our study employing DarkPatternDetector and exemplified by surveys conducted with Indian users, respondents encountered adaptive exploitation involving AI monetization propelling real time delays and evasive responses. This evidence highlights the importance of accessible user centered audit tools alongside enforceable design transparency. AI can be used for exploitation. DarkPatternDetector exemplifies how AI can also be used for accountability. Ethical algorithmic designs will ensure user roles are not confined as India continues constructing the most open digital infrastructure worldwide.